\RequirePackage{ifpdf}
\documentclass[12pt,a4paper,hyper]{JHEP3}

\usepackage{epsfig}
\usepackage{amsmath}
\usepackage{multirow}
\newcommand{\C}{{\mathbb C}}
\def\={\;  = \;}
\def\inn{\,\in\,}

\title{\center {Localization \& Exact Holography}}

\preprint{}
\author{
Atish Dabholkar$^{1}$, Jo\~ao Gomes$^{1}$, and Sameer Murthy$^{2}$\\

\it $^1${Laboratoire de Physique Th\'eorique et Hautes Energies (LPTHE)\\
\it{Universit\'e Pierre et Marie Curie-Paris 6; CNRS UMR 7589}\\
\it{Tour 13-14, 5$^{\grave{e}me}$ \'etage, Boite 126, 4 Place Jussieu} \\
\it {75252 Paris Cedex 05, France}}\\

\it $^2$Institute for Theoretical Physics and Spinoza Institute\\
\it University Utrecht, Leuvenlaan 4\\
\it 3584 CC Utrecht, The Netherlands \\

{\rm Emails : atish at lpthe.jussieu.fr, {gomes at lpthe.jussieu.fr},
{S.V.Murthy at uu.nl}}\\
}

\abstract{We consider the $AdS_{2}/CFT_{1}$ holographic correspondence near the horizon of big four-dimensional black holes preserving four supersymmetries in  toroidally compactified Type-II string theory. The boundary partition function of  $CFT_{1}$ is given by the known quantum degeneracies of these black holes. The bulk partition function is given by a functional integral over string fields in $AdS_{2}$.  Using recent results  on localization we  reduce the infinite-dimensional functional integral  to a finite number of  ordinary integrals over a space of  localizing instantons.  Under reasonable assumptions about the relevant terms in the effective action, these integrals can be evaluated  exactly to obtain a bulk partition function. It precisely reproduces  all terms in the exact Rademacher expansion  of the boundary partition function  as nontrivial functions of charges  except for  the Kloosterman sum which can in principle follow from an analysis of phases in  the background of orbifolded 
instantons. Our results can be regarded as a step towards proving `exact holography' in that the bulk and boundary partition functions computed independently agree for finite charges.  Since the bulk partition function defines the quantum entropy of the black hole,  our results enable the evaluation  of perturbative as well as nonperturbative quantum corrections to the Bekenstein-Hawking-Wald entropy of these black holes.}

\keywords{black holes, superstrings, holography}

\newenvironment{myenumerate}{
\begin{enumerate}
   \setlength{\itemsep}{1pt}
   \setlength{\parskip}{0pt}
   \setlength{\parsep}{0pt}}{\end{enumerate}}
\newenvironment{myitemize}{
\begin{itemize}
   \setlength{\itemsep}{1pt}
   \setlength{\parskip}{0pt}
   \setlength{\parsep}{0pt}}{\end{itemize}}

\renewcommand{\Im}{\mbox{Im}}

\newcommand{\IR}{\mathbb{R}}
\newcommand{\IZ}{\mathbb{Z}}

\newcommand{\Tr}{\mbox{Tr}}

\newcommand{\bem}{\begin{pmatrix}}
\newcommand{\eem}{\end{pmatrix}}

\def\vth{\vartheta}

\def\r{\rho}

\def\s{\sigma}
\def\g{\gamma}
\def\t{\tau}
\def\a{\alpha}

\def\m{\mu}
\def\n{\nu}

\def\h{\eta}
\def\l{{\lambda}}

\def\O{{\Omega}}
\def\D{\Delta}

\def\CF{{\cal F}}
\def\CS{{\cal S}}
\def\CL{{\cal L}}

\def\CN{{\cal N}}

\def\half{{\frac12}}

\def\CN{{\cal N}}

\def\bea{\begin{eqnarray}}
\def\eea{\end{eqnarray}}
\def\be{\begin{equation}}
\def\ee{\end{equation}}
\def\ba{\begin{align}}
\def\ea{\end{align}}
\def\bse{\begin{subequations}}
\def\ese{\end{subequations}}
\def\1F1{{}_1\!F_1}
\def\2F0{{}_2\!F_0}

\def\v{\varphi}

\def\a{\alpha}

\def\h3{$\textrm{H}_3^+$}

\def\IR{{\mathbb R}}
\def\IZ{{\mathbb Z}}

\newcommand{\beq}{\begin{equation}}
\newcommand{\eeq}{\end{equation}}
\newcommand{\ber}{\begin{eqnarray}}
\newcommand{\eer}{\end{eqnarray}}

\def\be{\begin{eqnarray}}
\def\ee{\end{eqnarray}}

\newcommand{\cO}{{\cal O}}
\newcommand{\cA}{{\cal A}}

\def\wh{\widehat}

\def\mod{{\rm mod}}

\def\CN{{\cal N}}

\def\CF{{\cal F}}

\def\CL{{\cal L}}

\def\CS{{\cal S}}

\def\CS{{\cal S }}

\def\Tr{{\rm Tr}}

\font\manual=manfnt
\def\dbend{\lower3.5pt\hbox{\manual\char127}}

\def\bar{\overline}
\def\CS{{\cal S}}
\def\CN{{\cal N}}

\def\rt2{\sqrt{2}}
\def\irt2{{1\over\sqrt{2}}}

\def\wt{\widetilde}

\def\s{\sigma}

\def\a{\alpha}

\def\g{\gamma}

\def\mod{{\rm mod}}

\font\cmss=cmss10
\font\cmsss=cmss10 at 7pt

\def\IL{\relax{\rm I\kern-.18em L}}
\def\IH{\relax{\rm I\kern-.18em H}}
\def\rlx{\relax\leavevmode}
\def\ZZ{\rlx\leavevmode\ifmmode\mathchoice{\hbox{\cmss Z\kern-.4em Z}}
 {\hbox{\cmss Z\kern-.4em Z}}{\lower.9pt\hbox{\cmsss Z\kern-.36em Z}}
 {\lower1.2pt\hbox{\cmsss Z\kern-.36em Z}}\else{\cmss Z\kern-.4em
 Z}\fi}

\def\Tr{{\rm Tr}}

\begin{document}

\section{Introduction}

In any consistent quantum theory of gravity such as string theory,  it should be possible to view a black hole as a statistical ensemble of quantum states.  This implies an extremely stringent  theoretical constraint on the theory that  the \textit{exact}  statistical  entropy of this ensemble must equal an appropriately defined quantum entropy of the black hole. 
Such a  constraint is also \textit{universal} in that  it must hold  in any  `phase' or compactification of the theory that admits a black hole.  It is therefore a  particularly useful guide in our explorations of  string theory in the absence of direct experimental guidance, especially given the fact that we do not know which phase of the  theory might describe the real world.  

The notion of  exact statistical entropy is \textit{a priori}  well-defined  as the logarithm of the dimension of the quantum Hilbert subspace corresponding to the ensemble. The notion of  exact quantum entropy of a black hole is more subtle but should be definable  as a quantum generalization of the Bekenstein-Hawking-Wald entropy \cite{Bekenstein:1973ur, Hawking:1974sw, Wald:1993nt, Iyer:1994ys, Jacobson:1994qe}.  A definition has recently been proposed  by Sen  \cite{Sen:2008yk, Sen:2008vm} for extremal black holes using holography in the  two dimensional anti de Sitter ($AdS_{2}$)
background near the horizon of the black hole. For a black hole of charge vector $(q, p)$, its quantum entropy  is defined as the logarithm  of the expectation value  $W(q, p)$ of a Wilson line  inserted on the boundary of the Euclidean $AdS_{2}$ space.

This definition expresses the exact quantum entropy as a formal functional integral over  all (spacetime) string fields in $AdS_{2}$. It is conceptually satisfying since it takes into account quantum effects from integrating over massless fields, keeps manifest all symmetries of the theory, 
and  reduces to the Wald entropy in the appropriate limit.  At the same time, it is rather difficult to work with, unless  one can figure out an efficient way to compute the functional integral. For supersymmetric black holes,  it is possible to use localization  techniques 
\cite{Witten:1988ze, Witten:1991zz, Witten:1991mk, Schwarz:1995dg, Zaboronsky:1996qn}
 to simplify this  infinite-dimensional functional integral enormously and reduce it  to a finite number of ordinary integrals  \cite{Dabholkar:2010uh}. In the present work we apply these results in the concrete context   of supersymmetric black holes  preserving four supersymmetries in $\mathcal{N}=8$ supersymmetric compactifications of string theory to four spacetime dimensions.  Since the structure of  the $\CN=8$ theory is particularly simple, it enables us to analytically perform the ordinary integrals that remain after localization and  evaluate $W(q, p)$ even after including nonpertubative effects.  
The resulting $W(q, p)$ matches in remarkable details with the  quantum degeneracies $d(q, p)$ of these black holes that are known independently.
These results are interesting from two related perspectives.
\begin{myitemize}

\item Our  results could be viewed as a step towards `proving' holography in the context of  $AdS_{2}/CFT_{1}$ correspondence.  Holography \cite{'tHooft:1993gx, Susskind:1994vu} has emerged as  one of the  central concepts concerning the microscopic degrees of freedom of quantum gravity. The heuristic principle that the  degrees of freedom of a quantum theory of gravity must scale with area rather than with volume has found its most precise realization in the $AdS_{d+1}/CFT_{d}$ correspondence \cite{Maldacena:1997re, Gubser:1998bc, Witten:1998qj}. Given the fundamental significance of the concept of holography, it is desirable to seek simple examples as we consider in this paper where  it might be possible to prove such an equivalence. We will compute the  partition functions  independently both in the bulk and the boundary for  arbitrary finite charges.

\item Our results could be viewed as the computation of  finite size quantum corrections to the leading Bekenstein-Hawking entropy of a black hole.  The  Bekenstein-Hawking entropy formula  is valid in the limit of large horizon area or large charges. Since it follows from the two-derivative Einstein-Hilbert action, it is independent of the `phase' or the compactification under consideration. By contrast, the finite size corrections  depend sensitively on the phase and contain a wealth of information about the details of compactification, the structure of higher-derivative effective action,  as well as  the spectrum of nonperturbative states in the theory. They are therefore very interesting as  a sensitive probe of the microscopic structure of the theory. 
\end{myitemize}

With these motivations we first  review the quantum entropy formalism and its relation with the $AdS_{2}/CFT_{1}$ correspondence in \S\ref{QE}.  In \S\ref{Micro} we describe the supersymmetric microstates in the $\CN=8$ theory  and their  degeneracies.   In   \S\ref{Sugra} we review the results of \cite{Dabholkar:2010uh} on the localization of the resulting functional integral and describe the explicit evaluation of the localizing integrals for the system under consideration in  \S\ref{Macro}  and conclude with remarks on open problems in \S\ref{Open}.

\section{Quantum Entropy and Holography  \label{QE}}

We now review the definition of  the quantum entropy  \cite{Sen:2008yk, Sen:2008vm} in the framework of 
  $AdS_{2}/CFT_{1}$ holographic correspondence in the near-horizon region of the black hole. 
In general, $AdS_{d+1}/CFT_{d}$
correspondence is obtained by focusing onto the  near horizon degrees of freedom in the spacetime around a  ($d$-1)-dimensional extremal black brane.  The dual $CFT_{d}$ is obtained by focusing onto low-energy excitations in the world-volume theory of the brane configuration. 
Quantum gravity in the  near horizon $AdS_{d +1}$ geometry is then expected to be equivalent to the quantum field theory of these low-energy excitations \cite{Maldacena:1997re, Gubser:1998bc, Witten:1998qj}.  

In our case, we have an extremal black 0-brane or  a  black  hole with electric charges $\{q_{I}\}$ and magnetic charges $\{ p^{I} \}$. The near horizon geometry   is  $AdS_{2} \times S^{2} \times T^{6}$. One can regard this as a compactification on $S^{2} \times T^{6}$ to obtain an effective theory on $AdS_{2}$ with an infinite number of fields. Magnetic charges are given by fluxes on $S^{2}$ of this compactification. The massless bosonic sector  contains the  metric, gauge fields 
$A^{I}$ with field strengths $F^{I}$, and scalar  fields $X^{I}$.   Classically, the metric on the Euclidean $AdS_{2}$ factor is
\be\label{metric}
 ds^2=  v_{*} \left[(r^2-1)d\theta^2+\frac{dr^2}{r^2-1}\right]   \, \qquad 1 \leq r < \infty; \, \qquad 0 \leq \theta < 2\pi \ . 
\ee
The scale $v_{*}$ of the horizon  as well as the values of the scalar fields $X_{*}^{I}$ and the electric fields $e^{I}_{*}$ at the horizon are determined  in terms of the charges $(q,  p)$ by the attractor mechanism.

Quantum mechanically, the $AdS_{2}$  functional integral is defined by summing over all field
configurations which asymptote to the these attractor values with the fall-off conditions \cite{Sen:2008yk, Sen:2008vm, Castro:2008ms}
\bea\label{asympcond}
d s^2 &=& v_{*} \left[ 
\left(r^2+\cO(1)\right) d\theta^2+ \frac{dr^2}{r^2+\cO(1)}  \right]\  .\\
X^{I} &= &X^{I}_{*} + \cO(1/r)\ ,\qquad
A^I = -i  \, e_{*}^{I} (r -\cO(1) ) d\theta\ .
\eea
All massive fields asymptote to zero because of their mass term. 

The functional integral for the partition function would be weighted by the exponential of the classical action given by a  Wilsonian effective action at some scale such as the string scale.
To make the classical variational problem well-defined,   it is necessary to add a boundary term 
\be
-i q^{I} \int A_{I}
\ee
to the action to   cancel the boundary terms arising from the variation of the bulk action for the gauge field.
 With this boundary term, the quantum bulk partition can be naturally interpreted as  an expectation value of a Wilson line inserted at the boundary
\be\label{qef}
W (q, p) = \left\langle \exp[-i \, q_I \oint_{\theta}  A^I]  \right\rangle_{\rm{AdS}_2}^{finite}\ ,
\ee 
where the superscript refers to a finite piece obtained by a procedure that we describe below.

The functional integral  \eqref{qef} has a well-known divergence as a consequence of the  infinite volume of $AdS_{2}$. This  can be removed by regularization and holographic renormalization.  We introduce a cutoff at  $r = r_{0}$ for a large $r_{0}$ to regularize the action.  The proper length of the boundary scales as $2 \pi \sqrt{v_{*}} r_{0}$.   Since the classical action    is an integral of a \textit{local} Lagrangian, it  scales as ${C_1 r_0 + C_0 + \cO(r_0^{-1})}$. The linearly divergent part can now be renormalized away  by a boundary counter-term which basically sets the origin of boundary energy. After this renormalization we can take  the cut-off to infinity to obtain a finite functional integral  weighted by the exponential of the  finite piece $C_0$.  We refer to $C_{0}$  as the renormalized action $S_{ren}$ which  is a functional of all fields and contains arbitrary higher-derivative terms\footnote{Regularizations corresponding with more general
cut-offs lead to the same renormalized action \cite{Sen:2009vz}.}.

It is worth emphasizing two peculiarities  of  $AdS_{2}$ that are significant  \cite{Sen:2008yk, Sen:2008vm} in this context. 
\begin{myitemize}
\item
For $d >2$,  the constant mode of the gauge field corresponding to the electric potential 
is dominant near the boundary and is hence kept fixed, while the $r$-dependent mode corresponding to
the electric field falls off at the boundary, and hence is allowed to fluctuate in the quantum theory. 
This corresponds to the grand-canonical ensemble where the chemical potential is held fixed\footnote{This is also true in  $d=2$ where the analysis is more subtle because of the Chern-Simons terms in $AdS_{3}$ 
\cite{Elitzur:1989nr, Dijkgraaf:2000fq}.}.
For $d=1$, the $r$-dependent mode of the gauge field corresponding 
to the electric field  grows linearly  and must be kept fixed, while the constant mode is
allowed to fluctuate. Fixing electric fields fixes all charges by Gauss law. This corresponds to the  microcanonical ensemble. 

\item
For  $d > 1$,  the $CFT_{d}$ of massless fields obtained by focusing on modes below a mass gap in the worldvolume  still allows for  a continuum of long wavelength, low energy excitations. 
For  $d=1$,  there are no spatial directions. The  boundary $CFT_{1}$ obtained by taking a low energy limit simply consists of the 
ground states in the charge sector $(q, p)$ and  has a degenerate and  finite-dimensional Hilbert space  with zero Hamiltonian.
The partition function of the  $CFT_{1}$ is then simply the \textit{number} $d(q, p)$  of these states.
Put another way, for any general $d$, conformal invariance allows all excitations with traceless stress tensor. In the special case of $d=1$, traceless stress tensor implies that the Hamiltonian is zero and there is no dynamics. This is consistent with the fact that  Lorentzian $AdS_{2}$  cannot support any finite energy fluctuations without disturbing the asymptotic boundary conditions  because of the large gravitational backreaction in low dimensions.

\end{myitemize}
The $AdS_{2}/CFT_{1}$ correspondence thus provides us  with a  satisfactory definition of quantum entropy as well as  a simple and yet nontrivial example of holography. It implies
\be
d(q, p) = W(q, p)
\ee
 The main challenge in the subsequent sections will be to find a context  where these formal definitions can be used for concrete calculations to compute both sides of  this equation.

\section{Microscopic Quantum Partition Function \label{Micro}}

Consider Type-II string compactified  on a 6-torus $T^{6}$. The resulting four-dimensional theory has $\CN=8$ supersymmetry with $28$ massless $U(1)$ gauge fields. A charged state is therefore characterized by $28$ electric and $28$ magnetic charges which combine into the $\bf 56$ representation of the U-duality group $E_{7, 7} (\mathbb{Z})$.  Under the  $SO(6, 6; \mathbb{Z})$ T-duality group, the $28$ gauge fields decompose as
\be
\textbf{28} =\textbf{12} + \textbf{16}
\ee
where the fields in the vector representation $\bf 12$  come from the NS-NS sector, while the fields in the spinor representation $\bf 16$  come from the R-R sector. We obtain an $\CN=4$ reduction of this theory by dropping four gravitini multiplets. Since each graivitini multiplet of $\CN=4$ contains four gauge fields, this amounts to dropping sixteen gauge fields which we take to be the R-R fields in the above decomposition.The  U-duality group of the reduced theory is
\be
SO(6, 6; \mathbb{Z}) \times SL(2, \mathbb{Z}) \,
\ee
where  $SL(2, \mathbb{Z})$ is the electric-magnetic S-duality group. 

\subsection{Charge Configuration \label{Config}}

We will be interested in one-eighth BPS dyonic states in this theory which perserve four of the thirty-two supersymmetries. 
To  simplify things, we consider the 6-torus to be the product  $T^4 \times S^{1} \times \wt S^{1}$ of a 4-torus and  two circles. Let  $n$ and $w$ be  the momentum and winding along the circle $S^{1}$, and  $K$ and $W$ be the corresponding Kaluza-Klein monopole and NS5-brane charges. Let $\wt n, \wt w, \wt K, \wt W$ be the corresponding charges associated with the circle $\wt S^{1}$. A general charge vector with these charges can be written as a doublet of $SL(2, \mathbb{Z})$
\begin{equation}\label{hetcharges}
    \Gamma = \left[
               \begin{array}{c}
                 Q \\
                 P \\
               \end{array}
             \right] =
             \left[
               \begin{array}{cccc}
                 {\wt n}& n  & {\wt w}& w \\
                 {\wt W}& W & {\wt K} & K \\
               \end{array}
             \right]_{B'},
\end{equation}
where the subscript $B'$ denotes a particular Type-IIB duality frame. The  T-duality invariants for this configuration are \cite{Giveon:1994fu}
\be
Q^{2} = 2(n w + \wt n \wt w) \, , \qquad P^{2} = 2(KW + \wt K \wt W) \, , \qquad Q \cdot P = n K + \wt n \wt K  + w W + \wt w \wt W \, ,
\ee
and the quartic U-duality invariant can be written as
\be
\Delta = Q^{2} P^{2} - (Q \cdot P )^{2} \, .
\ee
For our purposes it will suffice to excite only five charges 
\begin{equation}\label{charges}
    \Gamma =
             \left[
               \begin{array}{cccc}
                0& n  & 0& w \\
                 {\wt W}& W & {\wt K} & 0 \\
               \end{array}
             \right]_{B'}
\end{equation}
so that the T-duality invariants  are all nonzero. 
There are three other duality frames  that are of interest.

\begin{myitemize}
\item Frame $B$: In this frame the charge configuration becomes
\begin{equation}\label{chargesB}
    \Gamma =
             \left[
               \begin{array}{cccc}
                0& n  & 0& \wt K \\
                 Q_{1}& \wt n & Q_{5} & 0 \\
               \end{array}
             \right]_{B} \, ,
\end{equation}
where $Q_{1}$ is the number of D1-branes wrapping $S^{1}$ and $Q_{5}$ is the number of D5-branes wrapping $T^{4}\times S^{1}$. This frame is particularly useful for the microscopic derivation of the degeneracies described in 
\S\ref{microcount}. 
With $\wt K =1$, the Kaluza-Klein monopole interpolates between $\IR^{3} \times \wt S^{1}$  at asymptotic infinity and $\IR^{4}$ at the center. The momentum $\wt n$ at infinity becomes angular momentum at the center. This allows for a 4d-5d lift  \cite{Gaiotto:2005gf, David:2006yn} to relate the degeneracies of the four-dimensional state  to those of five-dimentional D1-D5 system carrying momentum $n$ and angular momentum $\wt n$.
\item Frame $A$: In this  frame the charge configuration becomes 
\begin{equation}\label{chargesA}
    \Gamma =
             \left[
               \begin{array}{cccc}
                0& q_{0}  & 0 & -p^{1} \\
                 p^{2}& q_{2}& p^{3} & 0 \\
               \end{array}
             \right]_{A} \, ,
\end{equation}
where $q_{0}$ is the number of D0-branes, $q_{2}$ is the number of D2-branes wrapping $S^{1}\times \wt S^{1}$, $p^{1}$  is a D4-brane wrapping $T^{4}$, $p^{2}$ is a D4-brane wrapping $\Sigma_{67}\times S^{1}\times \wt S^{1}$ and $p^{3}$ is a D4-brane wrapping $\Sigma_{89}\times S^{1}\times \wt S^{1}$ where $\Sigma_{ij}$ is a 2-cycle in $T^{4}$ along the directions $ij$. We will use this frame for localization in \S\ref{Sugra} and \S\ref{Macro}.
\item Frame $B^{''}$: In this frame the charge configuration becomes
\begin{equation}\label{chargesBdp}
    \Gamma =
             \left[
               \begin{array}{cccc}
                0& n  & 0 & Q_{5} \\
                Q_{3}& Q_{1}& Q_{3} & 0 \\
               \end{array}
             \right]_{B''} \, ,
\end{equation}
where all D-branes wrap the circle $S^{1}$ and an appropriate cycle in the $T^{4}$. 
\end{myitemize}

We can choose a charge configuration which is even simpler: 
\begin{equation}\label{chargesF}
    \Gamma =
             \left[
               \begin{array}{cccc}
                0& n  & 0& 1 \\
                1 & \nu & 1 & 0 \\
               \end{array}
             \right]
\end{equation}
where $n$ is a positive integer and $\nu$ takes values $0$ or $1$. 
The U-duality invariant is 
\be \label{udualinvt}
\Delta  = 4n - \nu^{2} \, .
\ee
It is clear that $\nu = \Delta$ modulo $2$, and so these states are completely specified by $\Delta$.
The states preserve four of the thirty-two supersymmetries. We will henceforth denote the degeneracies of these one-eighth BPS-states with charges \eqref{chargesF} by
$d(\Delta)$ instead of  $d(q, p)$. 

We should emphasize that a large class of states with the same value of $\Delta$ can be mapped by U-duality to  the state \eqref{chargesF} considered here but that does not exhaust all states. 
Note that the invariant $\Delta$ is the unique quartic invariant of the continous duality group $E_{7, 7}(\mathbb{R})$ but in general there are additional arithmetic duality invariants of the arithmetic group $G(\IZ)$  that cannot be written as invariants of $G({\IR})$.   As a result,  not all states with the same value of $\Delta$ are related by duality.  Classification of arithmetic invariants of $G(\IZ)$  is a subtle number-theoretic problem. For example, for the $\CN=4$ compactification where the duality group $O(22, 6; \IZ) \times SL(2, \IZ)$, essentially the only relevant arithmetic invariant  is given by $I = \gcd (Q \wedge P)$; and the degeneracies are known for all values of $I$ 
\cite{Banerjee:2007sr, Banerjee:2008ri, Banerjee:2008pu, Dabholkar:2008zy}. 
To our knowledge a similar complete classification  of  $E_{7, 7}(\mathbb{Z})$ invariants is not known at present. This would be a problem if one wishes to use canonical or a mixed ensemble. For our purposes, since we will working in the microcanonical ensemble, it will suffice to know the degeneracies for the states in the duality orbit of  \eqref{chargesF}.

\subsection{Microscopic Counting \label{microcount}}

The degeneracies of the 1/8-BPS dyonic states in the type II string theory on a $T^{6}$
are given in terms of the Fourier coefficients of the following counting function 
\cite{Maldacena:1999bp, Shih:2005qf, Sen:2008ta}: 
\bea\label{ourmicro}
F(\tau, z) &=&  \frac{\vth_1^2(\t, z)}{\eta^6(\t)} \, .
\eea  
where $\vartheta_{1}$ is  the Jacobi theta  function and  $\eta$ is the Dedekind function. With $ q:= e^{2\pi i \tau}$ and $ y:= e^{2\pi i z}$, they have the product representations
\bea \nonumber
\vth_1(\t, z) &=& q^{\frac{1}{8}} ( y^{\half} - y^{-\half})\prod_{n=1}^{\infty}(1 - q^{n}) (1 -  y q^{n}) (1 - y^{-1}q^{n}) \, ,\\ 
\eta(\tau) &=& q^{\frac{1}{24}} \prod_{n=1}^{\infty}(1 - q^{n}) \, \, .
\eea

The derivation of the counting function is simplest in the $B$ frame \eqref{chargesB} where we have  a D1-D5 system  in the  background of a single Kaluza-Klein monopole. By the 4d-5d lift, 
the momentum $\nu$ can be interpreted as 5d angular momentum.  The 
counting problem essentially reduces to counting bound states in \emph{five} dimensions of a single D1-brane bound to a single D5-brane carrying  $n$ units of momentum and $\nu$ units of angular momentum. Since the D1-brane can move inside the $D5$ anywhere on the $T^{4}$, the moduli space of this motion  is $T^{4}$. The function $F$ is  the  generalized elliptic genus of the corresponding superconformal field theory with target space $T^{4}$. This is  evident from the product representation which can be seen as coming from four bosons and four fermions.

Analysis of the  Fourier coefficients of $F$ simplifies enormously by the fact that $F$ is a \emph{weak Jacobi form}. 
We recall below  a few relevant facts about Jacobi forms \cite{Eichler:1985ja}. 
\begin{myenumerate}
\item \emph{Definition:}
A Jacobi form of  weight $k$ and index $m$  is a  holomorphic function $\varphi(\tau, z)$ from $\mathbb{H} \times\C$ to $\C$ which 
is ``modular in $\tau$ and elliptic in $z $'' in the sense that it transforms under the modular group as
  \be\label{modtransform}  \varphi(\frac{a\t+b}{c\t+d}, \frac{z}{c\t+d}) \ = 
   (c\t+d)^k \, e^{\frac{2\pi i m c z^2}{c\t +d}} \, \varphi(\t,z)  \qquad \forall \quad
   \left(\begin{array}{cc} a&b\\ c&d \end{array} \right) \in SL(2; \mathbb{Z}) \ee
and under the translations of $z$ by $\mathbb{Z} \tau + \mathbb{Z}$ as
  \be\label{elliptic}  \varphi(\t, z+\lambda\tau+\mu)\= e^{-2\pi i m(\lambda^2 \t + 2 \lambda z)} \varphi(\t, z)
  \qquad \forall \quad \l,\,\m \in \mathbb{Z} \, , \ee
where $k$ is an integer and $m$ is a positive integer.

\item \emph{Fourier expansion:}
Equations \eqref{modtransform} include the periodicities $\varphi(\t+1,z) = \varphi(\t,z)$ and $\varphi(\t,z+1) = \varphi(\t,z)$, so  $\varphi$ has a Fourier expansion
  \be\label{fourierjacobi} \varphi(\t,z) \= \sum_{n, r} c(n, r)\,q^n\,y^r\,, \qquad\qquad
   (q :=e^{2\pi i \t}, \; y := e^{2 \pi i z}) \ . \ee
Equation \eqref{elliptic} is then equivalent to the periodicity property
  \be\label{cnrprop}  c(n, r) \= C_{r}(4 n m - r^2) \ ,
  \qquad \mbox{where} \; C_{r}(D) \; \mbox{depends only on} \; r \, \mod\, 2m \ . \ee
The function  is called a \emph{weak} Jacobi form if it satisfies the condition
  \be\label{weakjacobi} c(n, r) \= 0\qquad   \textrm{unless}  \qquad n \geq 0 \, .\ee

\item \emph{Theta expansion:}
The transformation property (\ref{elliptic}) implies a 
Fourier expansion of the form
  \be\label{jacobi-Fourier} \v(\t, z) \= \sum_{\ell\inn \IZ} \;q^{\ell^2/4m}\;h_\ell(\t) \; e^{2\pi i\ell z} \ee
where $h_\ell(\tau)$ is periodic in $\ell$ with period $2m$.  In terms of the coefficients \eqref{cnrprop} we have
  \be\label{defhltau}  h_{\ell}(\t) \= \sum_{D} C_{\ell}(D) \,  q^{D/4m} \, \qquad \qquad (\ell \inn \IZ/2m \IZ)\;.  \ee
Because of the periodicity property of $h_{\ell}$, equation \eqref{jacobi-Fourier} can be rewritten in the form 
  \be\label{jacobi-theta} \v(\t,z) = \sum_{\ell\inn \IZ/2m\IZ} h_\ell(\t) \, \vartheta_{m,\ell}(\t, z)\,, \ee
where $\vartheta_{m,\ell}(\t,z)$ denotes the standard index $m$ theta function 
  \begin{eqnarray} \label{thetadef} \vartheta_{m,\ell}(\t, z) 
   \;:=\; \sum_{{\l\inn\IZ} \atop {\l\,=\,\ell\,(\mod\,2m)}} q^{\l^2/4m} \, y^\l \, \,
   \= \sum_{n \inn \mathbb{Z}} \,q^{m(n+ \ell/2m)^2} \,y^{\ell + 2mn}  \end{eqnarray}
This is the theta expansion of $\v$.  The vector $h := ( h_1, \ldots, h_{2m})$ transforms like a modular form of weight $k-\frac{1}{2}$ under $SL(2,\IZ)$.
\end{myenumerate}
With these definitions, $F(\t, z)$ is a weak Jacobi form of weight $-2$ and index $1$. The indexed degeneracies  for a state carrying $n$ units of momentum and $r$  units of angular momentum is then given by $c(n, r)$ in the Fourier expansion \eqref{fourierjacobi} of $F$. 
Using \eqref{cnrprop} 
for $m=1$,  we see that $c(n, r)$ depend only on $D = 4n-r^{2}$ and   $r$ mod $2$ which in this case  equals  $D$ mod $2$.  Hence, all information about the Fourier coefficients $c(n, r)$ of $F$ is contained in a single function  of $D$ alone which we denote by $C(D)$. Our task is thus reduced to determining $C(D)$  given \eqref{ourmicro}. 

To read off $C(D)$ more systematically we use the  theta expansion
\be\label{jacobi-theta2}  
F(\t,z)  = h_0(\t) \, \vartheta_{1,0}(\t, z)\, +  h_1(\t) \, \vartheta_{1,1}(\t, z)\,  .  
\ee
The functions $h_{\ell}(\tau)$ in this case are given explicitly by:
\bea \label{h0h1defs1}
h_{0} (\t) & = & - \frac{\vth_{1,1}(\t,0)}{\eta^{6}(\t)} = -2  -12 q - 56 q^{2}- 208 q^{3}\dots \\
\label{h0h1defs2} 
h_{1} (\t) & = & \frac{\vth_{1,0}(\t,0)}{\eta^{6}(\t)} =   q^{-\frac{1}{4}} \bigl(1 + 8 q + 39 q^{2} + \dots \bigr) 
\eea
For even and odd $D$, the coefficients $C(D)$ can  be read off   from these expansions of  $h_{0}$ and $h_{1}$ respectively using \eqref{defhltau}. 

It is clear that  $D$ can be identified with the duality invariant $\Delta$ in \eqref{udualinvt}. 
The degeneracies are then given  in terms of $ C(D)$ by
\be\label{Ctod}
d(\Delta) =  (-1)^{\Delta +1} C (\Delta) \, .
\ee
The factor of $(-1)^{\Delta}$ arises because  the state in five dimensional spacetime is fermionic for odd $\D$ and contributes to the index with a minus sign. The overall minus sign arises in relating the 4d degeneracies to the 5d degeneracies using the 4d-5d lift \cite{Shih:2005qf, Sen:2008ta}.

\subsection{Index, Degeneracy, and Fermions}\label{Index}

The first few terms in the Fourier expansion of $F$ are given by
\be
F(\t, z) = \frac{(y-1)^2}{y} \, - \, 2\,\frac{(y-1)^4}{y^2}\, q \, + \,  \frac{(y-1)^4(y^2-8y+1)}{y^3}\, q^2 \, + \, \cdots \ , 
\ee
In Table \eqref{tablefcoeffs} we tabulate the coefficients $C(\Delta)$ for the first few values  of $\Delta$. 
\begin{table}[h]   \caption{\small{Some Fourier coefficients}}   \vspace{8pt}    \centering
   \begin{tabular}{c|cccccccccccccc}   \hline   
       $\Delta$ &  -1 & 0 &3& 4& 7 & 8 &11& 12& 15 \\
       \hline $C(\Delta)$ & 1 & $-2$ &8 &$-12$&39&$-56$&152&$-208$&513\\   
   \hline    \end{tabular}   \label{tablefcoeffs}   \end{table}
   
It is striking that the sign of $C(\Delta) $ is alternating. This implies from \eqref{Ctod} that the degeneracies $d(\Delta)$ are always positive.  This is, in fact, true not only for the first leading coefficients but  for all Fourier coefficients, 
as can be seen from the equations  
\eqref{jacobi-theta2}--\eqref{h0h1defs2}. 
Mathematically, the alternating sign of the Fourier coefficients  is a somewhat nontrivial property of the specific Jacobi form \eqref{ourmicro} under consideration \cite{BringmannMurthy:2012}.  Physically, the positivity of $d(\Delta)$ is even more surprising. After all, these are  \emph{indexed} degeneracies  corresponding to a spacetime helicity supertrace for  a complicated bound states of branes. There is no \emph{a priori} microscopic reason why these should be all positive. 

Holography gives a simple physical explanation of the positivity \cite{Sen:2009vz, Sen:2010mz}. 
The near-horizon $AdS_{2}$ geometry has an $SU(1, 1)$ symmetry. If the black hole geometry leaves at least four supersymmetries unbroken, then closure of the supersymmetry algebra requires that the near horizon symmetry must contain the supergroup $SU(1, 1|2)$. This implies that that such a supersymmetric horizon must have $SU(2)$ symmetry which can be identified with spatial rotations.  If $J$ is a Cartan generator of this $SU(2)$, then for  a classical black hole with spherical symmetry, this could mean (depending on the ensemble) that either $J$  is zero  
or the chemical potential conjugate to $J$ is zero. As explained earlier, the $AdS_{2}$ path integral naturally fixes the charges and not the chemical potentials and hence $J=0$.  Together, this implies
\be\label{ind-deg}
\Tr (1) = \Tr (-1)^{J} \, ,
\ee
that is, index equals degeneracy and must be positive. 
For a more detailed discussion see \cite{Dabholkar:2010rm}.

Note the the index equals degeneracy only for the horizon degrees of freedom, but usually one  does not compute  the index of the horizon degrees of freedom directly. It is easier to compute the index  of the asymptotic states as a spacetime helicity supertrace which receives contribution also from the degrees of freedom external to the horizon. It is crucial that the  contribution of these external modes is removed from the helicity supertrace before checking the equality \eqref{ind-deg}.
Typically, modes localized outside the horizon come from fluctuations of supergravity fields and can carry NS-NS charges such as the momentum but not  D-brane charges \cite{Banerjee:2009uk,Jatkar:2009yd}.
In a given frame such as the $A$ frame where all charges come from D-branes, one expects that the Fourier coefficients of  $F(\t, z)$ will give the  degeneracies of only the horizon degrees of freedom. 

For  the Wilson line expectation value \eqref{qef} the equality \eqref{ind-deg} implies   that the functional integral with periodic boundary conditions for  the fermions must equal the functional integral with antiperiodic boundary conditions. This is possible for the following reason. All fermionic fields have nonzero $J$ and  couple to the Kaluza-Klein gauge field coming from the dimensional reduction on the $S^{2}$. As discussed above, the 
microcanonical boundary conditions \eqref{asympcond} for the functional integral instructs us to 
integrate over all the fluctuations of the constant mode. By a change of variables in the functional 
integral, one can change the origin of the constant mode of the gauge field, and therefore the 
periodic and antiperiodic boundary conditions for the fermionic fields are equivalent.


\subsection{Rademacher Expansion}

One can make very good estimates of  Fourier coefficients of a modular form using  an  expansion due to Hardy and Ramanujan. The leading term  of this expansion gives the Cardy formula. A generalization  due to 
Rademacher  \cite{Rademacher:1964ra} in fact  gives  an exact convergent  expansion for these coefficients in terms of the coefficients of the polar terms {\it i.e.} terms  with $ D<0$.

One can apply these methods to the Fourier coefficients of  the vector valued modular form $\{ h_{l} \}$ $(l = 0, \ldots 2m-1)$   of negative weight $-w$ to obtain  \cite{Dijkgraaf:2000fq, Manschot:2007ha}  a Rademacher expansion  for the coefficients  $C_{\ell}(D)$ \eqref{defhltau}
 \bea\label{radi} \nonumber
 C_\ell (D) &= & (2\pi)^{2-w} \sum_{c=1}^\infty 
  c^{w-2} \sum_{\wt\ell \inn \IZ/2m \IZ} \, \sum_{\wt D < 0} \, 
C_{\wt\ell}(\wt D) \,  
K(D,\ell,\wt D,\wt\ell;c) \, \left| \frac{\wt D}{4m} \right|^{1-w} \, \wt I_{1-w}
 \biggl[ {\pi\over c} \sqrt{| \wt D|  D}
\biggr] \, ,
\eea
%
%
where 
\begin{equation}\label{intrep}
 \wt{I}_{\rho}(z)=\frac{1}{2\pi
i}\int_{\epsilon-i\infty}^{\epsilon+i\infty} \, \frac{d\s}{\s^{\r +1}}\exp [{\s+\frac{z^2}{4\s}}]
\, 
\end{equation}
is called the modified Bessel function of index $\r$. This is 
related to the standard Bessel function of the first kind $I_{\rho}(z)$ by
\be
\wt I_{\rho}(z) = \big(\frac{z}{2} \big)^{-\rho} I_{\rho}(z) \, .
\ee
The sum over ($\wt\ell$, $\wt D$) picks up a contribution $C_{\wt\ell}(\wt D)$ from every  
non-zero term $q^{\wt D}$ with $\wt D < 0$ in $h_{\wt\ell}(\tau)$ \eqref{defhltau}.
The coefficients $ K\ell(D,\ell,\wt D,\wt\ell;c)$ are
generalized Kloosterman sums. For $c > 1$ it is  defined as
\be
\label{kloos0}
K(D,\ell;\wt D,\wt\ell;c):=
e^{-\pi i w/2}\sum_{{-c \leq d< 0}  \atop { (d,c)=1}}
e^{2\pi i \frac{d}{c} (D/4m)} \; M(\gamma_{c,d})^{-1}_{\ell\wt\ell} \; 
e^{2\pi i \frac{a}{c} (\wt D/4m)} \, , 
\ee
where
\be\label{gamform}
\gamma_{c,d} = \begin{pmatrix} a & (ad-1)/c \\ c & d \end{pmatrix}
\ee
is an element of $Sl(2,\IZ)$ and  $M(\gamma)$ is the  matrix representation  of $\gamma$ on the vector space spanned by the $\{ h_{l} \}$.  Note that it follows from \eqref{gamform} that $ad = 1\,  \mod\,  c$.

The Jacobi form $F(\t, z)$ has weight  $-2$ and index $m=1$, so its theta expansion gives a two-component vector $\{h_{0}, h_{1}\}$ of modular forms of weight $w = -5/2$. 
Since  there is  only a single polar term $(\wt\ell=1, \wt D=-1)$, the Rademacher expansion takes the form:
 \be\label{rademsp} 
 C(D) =   2{\pi} \, \big( \frac{\pi}{2} \big)^{7/2} \, \sum_{c=1}^\infty 
  c^{-9/2} \, K_{c}(D) \; \wt I_{7/2} \big(\frac{\pi \sqrt{D}}{c} \big)  \, , 
\ee
where the Kloosterman sum $K_{c} (D) $  is defined by
\bea
\label{kloos}
K_{c}(D) :=
e^{5\pi i /4}\sum_{-c \leq d< 0; \atop (d,c)=1}
e^{2\pi i \frac{d}{c} (D/4)} \; M(\gamma_{c,d})^{-1}_{\ell 1} \; 
e^{2\pi i \frac{a}{c} (-1/4)} \qquad  \,   \,    \qquad  
\eea
with $\ell = D \, \mod \, 2$ and $ad = 1 \, \mod \, c$.

Under the $SL(2,\IZ)$ generators, the modular form $h_{\ell}(\t)$ transform as
\bea
h_{0} (\tau + 1)  =  h_{0} (\t) \, , \quad  \;  &&  \qquad h_{0} (-1/\tau)  =  \frac{1+i}{2} \, \t^{-5/2}  \big(h_{0} (\t) + h_{1}(\t) \big) \, ; \\ 
h_{1} (\tau + 1)  = -i \, h_{1} (\t) \, , &&  \qquad h_{1} (-1/\tau) =  \frac{1+i}{2} \, \t^{-5/2}  \big(h_{0} (\t) - h_{1}(\t) \big) \, .
\eea
From these transformations, we can read off the matrices $M(\g)$ for the generators  $S$  and $T$ 
\be
T=\bem 1 & 1 \\ 0 & 1 \eem \, ,  \qquad S = \bem 0 & 1 \\ -1 & 0 \eem 
\ee 
to be
\be\label{MTMS}
M(T) = \bem 1&0\\0&-i \eem \, , \qquad M(S) = \frac{e^{\pi i/4}}{\sqrt{2}} \bem 1&1\\1&-1 \eem \, . 
\ee
Using the expression for a general $SL(2, \IZ)$ matrix $\g$ in terms of the generators $S$ and $T$, and the 
representation \eqref{MTMS},  we can obtain the representation $M(\g)$.

We see from \eqref{rademsp} that the microscopic degeneracy is an infinite sum of the form
\be\label{dexp}
d(\Delta)  = \sum_{c=1}^{\infty}d_{c} (\Delta) \, .
\ee
where each term is given by
\be\label{dc}
d_{c} (\Delta)  =  (-1)^{ \Delta +1} \, 2\pi \big(\frac{\pi}{ \Delta}\big)^{7/2} \, 
 I_{\frac{7}{2}}\big(\frac{\pi \sqrt{\Delta}}{c}\big) \, \frac{1}{c^{9/2}}  K_{c} (\Delta) \, .
\ee
It is easy to check that 
\be\label{K1}
K_{1} = (-1)^{ \Delta +1} \frac{1}{\sqrt{2}} \, .
\ee 
We will see that the Wilson line from the macroscopic side also naturally has the same expansion 
\be\label{Wexp}
W(\Delta)  = \sum_{c=1}^{\infty}W_{c} (\Delta) \, ,
\ee
coming from $\IZ_{c}$ orbifolds of $AdS_{2}$. Our objective then is to compute each of these terms exactly using localization. We compute the leading term $W_{1}(\Delta)$ in \S\ref{Macro} and the subleading terms corresponding to $c>1$ in \S\ref{Nonpert}. 

\section{Localization of  Functional Integral in Supergravity\label{Sugra}}

Evaluating the formal functional integral \eqref{qef} over string fields for  $W(q, p)$ is of course highly nontrivial.  To proceed further, we first integrate out the infinite tower of massive string modes and massive Kaluza-Klein modes to obtain a \emph{local} Wilsonian effective action for the massless supergravity 
fields keeping all higher derivative terms. We can  regard the ultraviolet finite string theory as providing a supersymmetric and consistent cutoff at the string scale. Our task is then reduced to evaluating a functional integral in supergravity.  The near horizon geometry preserves eight superconformal symmetries and the  action, measure, 
operator insertion, boundary conditions of the functional integral \eqref{qef} are all supersymmetric\footnote{Supersymmetry of the Wilson line and the action is discussed in the appendix of \cite{Dabholkar:2010uh}.}. The formal supersymmetry of the functional integral makes it possible to apply localization techniques \cite{Dabholkar:2010uh, Banerjee:2009af} to evaluate it.

To apply localization to our system, we drop two gravitini multiplets to obtain a $\CN=2$ theory and also drop the hypermultiplets to consider a reduced theory.  This theory contains a supergravity multiplet coupled to eight vector multiplets with a duality group
\be
SO(6, 2; \mathbb{Z}) \times SL(2, \mathbb{Z}) \, .
\ee 
In the effective action for these fields we will further ignore the D-type terms. For a partial justification for this reduction in this context and for further  discussion see \S\ref{Open} and  \cite{Dabholkar:2010uh}. 
We will denote  the functional integral \eqref{qef} restricted to this reduced theory by $\widehat W(q, p)$ which is what we compute in the subsequent sections. We  find that  $\widehat W(q, p)$ itself agrees perfectly with \eqref{rademsp} for $d(q, p)$. This rather nontrivial agreement  can be regarded as  post-facto evidence that the reduced theory correctly captures the relevant physics.

\subsection{Functional Integral in $\mathcal{N}=2 $ Off-shell Supergravity \label{Functional}}

Localization of the supergravity functional integral is considerably simplified in the off-shell formalism. 
The main advantage of  the off-shell formalism is that  the supersymmetry transformations are specified  once and for all, and do not need to be modified as one modifies the action with higher derivative terms. Consequently, the localizing instantons that we describe below do not depend upon the form of the physical action. 
The problem of finding the  model-independent localizing instantons is  then cleanly separated from the problem of  evaluating the renormalized action for a specific physical action.

In the  off-shell formalism for  $\mathcal{N}=2$ supergravity  developed in \cite{deWit:1979ug, deWit:1984px, deWit:1980tn} the vielbein and its superpartners reside in the Weyl multiplet. In addition, we  consider  $n_{v} +1$ vector multiplets with the field content
\be\label{Vectorfields}
{\bf X}^{I} = \left( X^{I}, \O_{i}^{I}, A_{\mu}^{I}, Y^{I}_{ij}  \right) \, , \quad I = 0,  \ldots, n_{v}\, .
\ee
For each $I$,  the multiplet contains eight bosonic and eight fermionic degrees of freedom: $X^{I}$ is a complex scalar, the gaugini $\O^{I}_{i}$ are an $SU(2)$ 
doublet of chiral fermions, $A^{I}_{\mu}$ is a vector field, and $Y^{I}_{ij}$ are an $SU(2)$ triplet of 
auxiliary scalars. The auxiliary fields $Y^{I}_{ij}$ play a very important role in localization.

Localization is a general technique for evaluating superintegrals of the form
\begin{equation}
I   = \int_{\mathcal{M}} d\mu  \, h \, e^{-\mathcal{S} } \ . 
\end{equation}
Here $\mathcal{M}$ is the supermanifold with integration measure $d\mu$, which has  
an odd (fermionic) vector field $Q$  which squares to a compact bosonic vector field $H$; $h$, $S$,  
and the measure are all invariant under $Q$. 
To evaluate this integral one first deforms it  to
\begin{equation}
I (\lambda)  = \int_{\mathcal{M}} d\mu  \, h \, e^{-\mathcal{S}  - \lambda QV} \ , 
\end{equation}
where $V$ is a fermionic, H-invariant function which means  $Q^{2} V = 0$ and  $Q V$ is Q-exact. One has 
\begin{equation}
\frac{d}{d\lambda}\int_{\mathcal{M}} d \mu   \, h  \, e^{- \mathcal{S} - \lambda QV} = \int_{\mathcal{M}} d \mu   \, h  \, QV \, e^{- \mathcal{S} - \lambda QV} = \int_{\mathcal{M}} d \mu   \, Q( h  \, e^{- \mathcal{S} - \lambda QV}) = 0 \ , 
\end{equation}
and hence $I(\lambda)$ is independent of $\lambda$. 
This implies that one can perform the integral $I(\lambda)$ for any value of $\lambda$ and in particular for $\lambda \rightarrow \infty$.  
In this limit, the functional integral  localizes onto the  critical points of the functional $S^{Q} := QV$ 
which we refer to as the localizing instanton solutions.  
One can choose in particular,
\begin{equation}\label{locV}
V = (Q\Psi, \Psi)
\end{equation}
 where $\Psi$ are the fermionic coordinates with some positive definite inner product  defined on the fermions.
In this case, the bosonic part of  $S^{Q}$ can be written as a perfect square $(Q\Psi, Q\Psi)$, and hence critical points of $S^{Q}$ are the same as the zeros of $Q$. 
 Let us denote the set of zeros of 
 $Q$ by $\mathcal{M}_{Q}$. 
The reasoning above shows that the integral over the supermanifold $\mathcal{M}$ localizes to an integral over the submanifold $\mathcal{M}_{Q}$. 
In the large $\lambda$ limit, the integration for directions transverse can be performed exactly in the saddle point evaluation. One is then left with an integral over the submanifold $\mathcal{M}_{Q}$
with a measure $d\mu_{Q}$ induced on the submanifold. 

In our case, $\mathcal{M}$ is the field space of off-shell supergravity, $\mathcal{S}$ is the off-shell supergravity action with appropriate boundary terms, $h$ is the supersymmetric Wilson line. 
To localize, we will choose the fermionic symmetry generated by   the supersymmetry generator $Q$ which squares to $4(L-J)$, where $L$ is the generator of rotations of the Poincar\'e disk 
and $J$ is the generator of rotations of $S^{2}$.
With this choice for $Q$,  the localizing Lagrangian is then defined by 
\begin{equation}\label{loclag}
\CL^{Q} := QV \quad {\rm with} \quad V := (Q \Psi, \Psi) \, ,
\end{equation}
where $\Psi$ refers to all fermions in the theory. The localizing action is then defined by
\begin{equation}
\mathcal{S}^{Q} = \int d^{4} x \sqrt{ g} \, \CL^{Q} \, .
\end{equation}
The localization equations that follow from this action  are
\begin{equation}
Q \Psi = 0 \, . 
\end{equation}
These are the equations that we needs  to solve subject to the 
$AdS_{2}$ boundary conditions. The scalar fields are fixed to their attractor values
\be\label{attract}
X^{I}_{*} = \frac{1}{2}(e_{*}^{I} + i p^{I})\,
\ee
where $e_{*}^{I}$ are the attractor value of the electric fields  determined in terms of the charges $(q, p)$.

The solution to this system of differential equations subject to the $AdS_{2}$ boundary condition turns out to be surprisingly simple and can be given in a closed form  \cite{Dabholkar:2010uh}.
 The most general solution  parametrized by $(n_{v} +1)$ real parameters $\{C^{I}\}, \ I = 1, \ldots, n_{v} +1,
$ and is given by the  field configurations
\begin{eqnarray} \label{HEKEsol}
 X^{I}  =  X^{I}_{*} +  \frac{C^{I}}{r} \ , \qquad  \bar X^{I}  =  \bar X^{I}_{*} +  \frac{C^{I}}{r} \ , \qquad Y^{1I}_{1} = - Y^{2I}_{2} =  \frac{2C^{I}}{r^{2}} \,\,  , \end{eqnarray}
with other fields fixed to their attractor values\footnote{It was shown in \cite{Dabholkar:2010uh} that the gauge fields in the vector multiplets are not excited for the localizing solutions. A similar analysis remains to be done to show that there are no other more general localizing solutions exciting fields in the supergravity multiplet. In what follows we will assume this to be true.}.  The real parameters 
$\{C^{I}\} $ can be thought of as the collective coordinates of the localizing instantons.  The functional integral 
of supergravity thus localizes onto a finite number of ordinary bosonic integrals over  $\{C^{I}\}$ which enormously simplifies the evaluation of the Wilson line \cite{Dabholkar:2010uh}. 
So far we have not assumed any particular form of the physical action. 
As emphasized in \cite{Dabholkar:2010uh}, these localizing instanton solutions are \textit{universal} in that they follow simply from the off-shell supersymmetry transformation laws of the vector multiplet fermions and hence are independent of the physical action. 

When the action contains only F-type terms, it is  governed by a single  prepotential $F(X^{I}, \hat A)$ which  is a meromorphic function of its arguments  and 
obeys the homogeneity condition:
\be \label{homogen}
F(\lambda X, \l^{2} \hat A) = \lambda^2
F( X, \hat A)\, .
\ee
where  $\hat A$ is an auxiliary field from the supergravity multiplet. Terms depending on $\hat A$ lead to  higher derivative terms in the action \cite{Mohaupt:2000mj}.   

To obtain the integrand over this localizing integral, one must substitute the solution \eqref{HEKEsol} into the physical action  and extract the finite part as a function of the collective coordinates $\{ C^{I }\}$   following the prescription in \S\ref{QE}. 
One obtains \cite{Dabholkar:2010uh} a remarkably simple expression for the the renormalized action for the localizing instantons as a function of the collective coordinates $\{ C^{I} \}$:
\be\label{Srenfinal}
\CS_{\rm ren} =  - \pi   q_I  e^I_* - 2 \pi  q_{I}  C^{I}  
 - 2 \pi i \big(F(X_{*}^{I} + C^{I}) -  \bar F(X_{*}^{I} + C^{I}) \big) \ . 
\ee 
Using the scalar attractor values  \eqref{attract}
and the new variable 
\be\label{ephi}
\phi^I := e_{*}^I+2 C^I \ ,
\ee
we can express the renormalized action as
\begin{eqnarray} \label{Sren}
 \mathcal{S}_{ren}(\phi, q, p) =  - \pi  q_I   \phi^I + \mathcal{F}(\phi, p)\, .
\end{eqnarray}
with
\begin{equation} \label{freeenergy2}
\mathcal{F}(\phi, p) = - 2\pi i \left[ F\Big(\frac{\phi^I+ip^I}{2} \Big) -
 \bar{F} \Big(\frac{\phi^I- ip^I}{2} \Big) \right] \, .
 \end{equation}
 Written this way, note that  the prepotential is evaluated precisely for values of the scalar fields at the origin of the  $AdS_{2}$ and not at the boundary of the  $AdS_{2}$.  At the boundary, the fields remain pinned to their attractor values  and in particular  the electric field remains fixed  as required by the microcanonical boundary conditions of the functional integral. The collective coordinates  $\phi^{I}$  in \eqref{ephi}  still fluctuate because  $C^{I}$ take values over the  real line. 
 
The renormalized action $S_{ren}(\phi)$ has the same  functional form as the classical entropy function. In particular, its extrema $\phi = \phi_{*}$ correspond to the attractor values of the scalar fields and its value at the extremum
 $S_{ren}(\phi^{*})$ equals the Wald entropy for the local Lagrangian described with a prepotential $\CF$.  However, the  physics behind  the renormalized action  is completely different. Unlike the classical entropy function which is essentially a classical on-shell object, the renormalized action is a quantum object  obtained after a complicated holographic renormalization procedure using an off-shell localizing field configuration \eqref{HEKEsol}. Even though the scalar  fields in the localizing solution  asymptote to the attractor values at the boundary of the $AdS_{2}$, they 
have a nontrivial coordinate dependence  in the bulk  and they take the value $ X^{I}_{*} +  C^{I}$ at the center of $AdS_{2}$. In particular, they are excited away from their attractor values and  are no longer at the minimum  of $S_{ren}$.  Even though the scalar fields thus `climb up the potential' away from the minimum of the entropy function, the localizing solution remains Q-supersymmetric (in the Euclidean theory) because the  auxiliary fields $Y^{I}_{ij}$ get
excited appropriately to satisfy the Killing spinor equations. This is what enables us to integrate over $\phi$ for values  in field space far away from the on-shell values.

 The infinite dimensional functional integral \eqref{qef}  for the Wilson line in the reduced theory can thus be written as a finite integral
 \begin{equation}\label{integral}
 \wh W (q, p) = \int_{\mathcal{M}_{Q}}   e^{ -\pi  \phi^{I} q_{I}}  \, e^{\mathcal{F}(\phi, p)}
 \,  \, |Z_{inst}|^{2} \, Z_{det}\,  [d\phi]_{\mu}
\end{equation}
The measure of integration $[d\phi]_{\mu}$ is computable from  the original measure $\mu$ of the functional integral of massless fields of string theory by standard collective coordinate methods. The factor $Z_{det}$ is the one-loop determinant of the quadratic fluctuation operator around the localizing instanton solution. Such one-loop determinant factors in closely related problems have been computed in \cite{Pestun:2007rz, Gomis:2009ir}. 
We have included $|Z_{inst}|^{2}$ to include possible contributions from brane instantons which is partially captured by the topological string for a class of branes.

Note that the exponential of the integrand is in the spirit of the conjecture by Ooguri, Strominger, and Vafa \cite{Ooguri:2004zv}. Our treatment differs from \cite{Ooguri:2004zv} in that the natural ensemble in our 
analysis is the microcanonical one. Moreover, we will be able determine the measure factor from first principles and the determine the subleading orbifolded localizing instantons that contribute to the functional integral. For earlier related work see \cite{Beasley:2006us, Denef:2007vg}.

To compute $\widehat W(q, p)$, it is necessary to evaluate all these factors explicitly and then  perform the finite dimensional integral over $\phi$. This is what we will do for our system in \S\ref{Macro}.  For the $\CN=2$ reduction of the $\CN=8$ theory that we consider, $n_{v}=7$ and the prepotential is given by
\be \label{ourprepot}
F(X) = -\frac{1}{2} \frac{X^{1}C_{ab}  X^{a} X^{b} }{X^{0}} \, , \qquad \qquad a, b = 2, \ldots, 7 \, .
\ee 
where $C_{ab}$ is the intersection matrix of the six 2-cycles of $T^{4}$. In  our normalization, it is given  by
\be
C_{ab} =  \left(\begin{array}{cc}0 & 1 \\1 & 0\end{array}\right) \otimes \textbf{1}_{3\times 3}
\ee
where $\textbf{1}_{3\times 3} $ is a $3 \times 3$ identity matrix. This prepotential describes the classical two-derivative supergravity action. Note that this does not depend the field $\hat A$ because there are no higher-derivative quantum corrections to the prepotential. 

\subsection{Integration Measure}

The measure $[d\phi]_{\mu}$ is inherited from the standard measure on field space in the original functional integral. 
The collective coordinates $\{\phi^{I}\}$ of the localizing instanton solutions correspond to the values of the scalar fields $\{ X^{I} \}$  at the center of the $AdS_{2}$. The functional integration measure for the scalar fields is a pointwise product of  integration measure over the scalar manifold. The metric and hence the measure on the scalar manifold can be read off from the kinetic term of the scalar fields \cite{Mohaupt:2000mj, LopesCardoso:2000qm}.
The scalar kinetic action is
\begin{equation}\label{scalarKE}
8\pi  \mathcal{L}=  \sqrt{|g|} g^{\mu\nu}\left[ i(\partial_{\mu}F_I+i\mathcal{A}_{\mu}F_I)(\partial^{\mu}\bar{X}^I-i\mathcal{A}^{\mu}\bar{X}^I)+h.c. \right] \, , 
\end{equation}
where $\mathcal{A}_{\mu}$ is the gauge field for the $U(1)$ gauge symmetry of the off-shell supergravity 
theory. This field does not have a kinetic term and it is therefore determined by its equation of motion to be
\begin{equation}
 \mathcal{A}_{\mu}^{*}=\frac{1}{2}\frac{\bar{F}_I\vec{\partial}_{\mu}X^I-\bar{X}^I\vec{\partial}_{\mu}F_I}{-i(\bar{F}_IX^I-F_I\bar{X}^I)} \, .
\end{equation}
The Lagrangian $ 8\pi  \mathcal{L}^*$ computed by substituting  $\mathcal{A}_{\mu}^{*}$ in \eqref{scalarKE} becomes
\begin{equation}\label{scalarKE1}
 - \sqrt{|g|} g^{\mu\nu}\left[N_{IJ}\partial_{\mu} X^I\partial_{\nu} \bar{X}^J - \frac{e^{-K}}{4}( K_{I}\partial_{\mu}X^I - \bar K_{ I}\partial_{\mu} \bar{X}^I) ( K_{I}\partial_{\nu} X^I -\bar K_{I}\partial_{\nu} \bar{X}^I) \right] \, ,
\end{equation}with
\begin{eqnarray}
\label{Ndef} N_{IJ}&:= & -i (F_{IJ} - \bar F_{IJ}) = 2\,\textrm{Im}(F_{IJ}) \, , \\
\label{Kdef} e^{-K}&:= & -i(X^I \bar{F}_I   - \bar{X}^I F_I) \, , \\
\label{Ldef} K_I &:= &  \frac{\partial K}{\partial X^{I}} =  i e^{K}\left( \bar F_I -  {F}_{IJ}\bar X^J \right)  .
\end{eqnarray}

The metric $g_{\mu \nu}$ is not the physical metric of Poincar\'e supergravity because it does not 
come with the canonical kinetic term.  It is related to the dilatation-invariant physical metric $G$ as 
\be\label{physmetric}
 G_{\mu\nu} =  e^{-K}g_{\mu\nu} \, ,
\ee
whose kinetic term is given by the standard Einstein-Hilbert action. We have 
\be\label{gGrel}
\sqrt{|g|} g^{\mu\nu} = e^{K} \sqrt{|G|} G^{\mu\nu} \, . 
\ee

It is natural to define the scalar functional integral measure using the physical metric $G_{\mu\nu}$. 
The measure can be determined by the metric induced by the inner product in field space:
\be
(\delta X, \delta X) = \int d^{4}x \,  \sqrt{|G|}\,  \delta X \, \delta X \, .
\ee
Substituting $X^{I} = (\phi^{I} + ip^{I} )/2$ in \eqref{scalarKE1}, and using  \eqref{physmetric}, \eqref{gGrel},
we obtain the induced metric on the localizing submanifold in the field space
\be\label{lineelem1}
d\Sigma^{2} =  M_{IJ} \, \delta \phi^{I} \delta \phi^{J} \, , 
\ee
with 
\be\label{Mmat1}
M_{IJ} = e^{K} \left[ N_{IJ} -\frac{e^{K}}{4} (K_{I} - \bar K_{I} )   (K_{J} - \bar K_{J} ) \right] \, .
\ee

It  is possible to write  the metric on the localizing manifold entirely
in terms of the K\"ahler potential\footnote{Upon gauge-fixing, on the space of projective coordinates, $K$ becomes the K\"ahler potential.  We will refer to  $K$ as the K\"ahler potential even though we do not fix any gauge here.}
 $K$ \eqref{Kdef}. 
It is easy to check that
\begin{eqnarray} \label{LNiden}
N_{IJ}& = & - \frac{\partial^{2} e^{-K}}{\partial X^{I} \bar X^{J}}=  
e^{-K}\left( \frac{\partial^{2} K}{\partial X^{I} \partial \bar X^{J}} -  \frac{\partial K}{\partial X^{I}}  \frac{\partial K}{\partial \bar X^{J}} \right) \, . 
\end{eqnarray}
Defining the metric $K_{IJ}$ in terms of the K\"ahler potential in the usual way 
\be\label{kahmet}
K_{I J} := \frac{\partial^{2 } K}{\partial X^{I} \partial \bar X^{J}} \, , 
\ee
and using \eqref{LNiden}, we can write the Lagrangian \eqref{scalarKE1} entirely in terms of 
the K\"ahler potential:
\begin{equation} \label{scalarKE2}
 8\pi  \mathcal{L}= - \sqrt{|g|} g^{\mu\nu}e^{-K}\left[ K_{IJ}\partial_{\mu} X^I\partial_{ \nu} \bar{X}^J - \frac{1}{4} \partial_{\mu}K \partial_{\nu} K \right] \, .
\end{equation}

Substituting  $X^{I} = (\phi^{I} + ip^{I} )/2$ in \eqref{scalarKE2}, we can rewrite the moduli space metric 
\eqref{lineelem1} as 
\be\label{Mmat2}
M_{IJ } = K_{IJ} -\frac{1}{4} \frac{\partial K}{\partial \phi^{I}} \frac{\partial K}{\partial \phi^{J}} \, . 
\ee
Since the metric $K_{IJ}$ is given in  terms of the K\"ahler potential \eqref{kahmet}, this expresses the moduli space metric $M_{IJ}$ entirely in terms of the K\"ahler potential. 
The measure on the localizing manifold is simply the measure induced by this metric and is given by 
\be
\prod_{I =0}^{n_{v}} d\phi^{I} \sqrt{\det (M)} \, .
\ee

\section{Macroscopic Quantum Partition Function \label{Macro}}

The two-derivative action of $\CN=8$ is invariant under the continuous duality group $E_{7, 7}(\mathbb{R})$. We therefore expect to be able to write the macroscopic answer in terms of  $\Delta$  which  is the unique quartic invariant of  $E_{7, 7}(\mathbb{R})$. For this purpose, we will first write the renormalized action in new variables   so that it depends only on the invariant $\Delta$ and then work out the measure in the same variables to obtain a manifestly duality invariant expression for the Wilson line.

\subsection{Renormalized Action and Duality-invariant Variables \label{renaction} }

As discussed in \S\ref{Config} the electric and magnetic charge vectors  $Q$ and $P$ respectively  are related to the charges in the Type-IIA frame \eqref{chargesA} by
\bea
Q  =  ( q_{0}, -p^{1}; q_{a}) \, \qquad
P  = (q_{1}, p^{0}; p^{a} ) \quad .
\eea
The inner product is defined for example by
\be 
P \cdot P = 2  \, q^{1}  p^{0} +  p^{a} \, C_{ab} \, p^{b} \, ,
\ee
The charge configuration \eqref{chargesF} has only five nonzero charges $q_{0} = n $, $q_{1} = l $, $p^{1} = -w$, and $p^{2}$, $p^{3}$.  Hence, the three T-dualiy invariants  all have nonzero values given by
\be\label{tinvt}
Q^{2 } = 2 \, n  w \, , \quad P^{2} = 2 \, p^{2}  p^{3}\, , \quad Q \cdot P = w \, l  \, .
\ee

The natural variables to start with are the projective coordinates
\begin{equation}\label{para1}
S := X^{1}/X^{0} \, , \quad   \quad T^{a} := X^{a}/X^{0} \quad a = 2, \ldots, n_{v} \, ,
\end{equation}
with real and imaginary parts defined by
\be\label{para2}
S:=  a  + is \, , \quad T^{a} :=  t^{a} + i r^{a} \, .
\ee
For our localizing instanton solutions we obtain
\bea
a = \phi^{1}/\phi^{0} \, ,&\quad& s = -w/\phi^{0} \\
t^{a} = \phi^{a}/\phi^{0} \, , &\quad& r^{a} = p^{a}/\phi^{0}  \, . 
\eea
The renormalized action \eqref{Srenfinal} for this charge configuration and prepotential \eqref{ourprepot} is 
\be\label{Srenour}
S_{ren} = -\frac{\pi}{2\phi^{0}} \left[ -w (\phi^{2} - P^{2}) + 2 \, \phi^{1} (\phi \cdot P) \right]
-\pi  n  \phi^{0} - \pi  l  \phi^{1} \, ,
\ee
where $\phi^{2}= \phi^{a} \, C_{ab} \, \phi^{b}$ and $\phi \cdot P = \phi^{a} \, C_{ab} \, P^{b}$.
Using the parametrization \eqref{para1} and \eqref{para2} and the T-duality invariants \eqref{tinvt} it can be written as
\be\label{Sren1}
S_{ren} = \frac{\pi}{2} \left[P^{2} s + \frac{Q^{2}}{s} + \frac{2 \, Q\cdot P \, a}{s}\right] 
-\frac{\pi w^{2} \, t^{2}}{2s} + \frac{\pi aw \, t \cdot P}{s}  \, \, .
\ee

Our  next goal will be to define integration variables to write the action entirely in terms of the U-duality invariant  $\Delta$.
Since the action is quadratic in the $t^{a}$ variables, it is useful to complete the squares by defining
\be
\tau^{a} = \frac{w}{\sqrt{s}} \left( t^{a} - \frac{a \, p^{a}}{w}  \right)
\ee 
so that 
\be\label{Sren2}
S_{ren} = \frac{\pi}{2} \left[P^{2} s + \frac{Q^{2}}{s} + \frac{P^{2} \, a^{2}}{s} + \frac{2 \, Q\cdot P \, a}{s}\right] 
-\frac{\pi \, \tau^{2}}{2}  \, \, .
\ee
Note that the parenthesis  is a manifestly S-duality invariant combination which  is quadratic in the axion variable $a$. So we complete the square again by defining
\be\label{defbig}
\sigma = \frac{\pi P^{2 }s}{2} \, , \quad 
\alpha = \frac{1}{\sqrt{\sigma}} \left( P^{2}a  + Q\cdot P \right)
\ee
The renormalized action then becomes
\be
S_{ren} =  \left( \sigma + \frac{z^{2}}{4\sigma} \right) -\frac{\pi \, \tau^{2}}{2}  + \frac{\pi \, \alpha^{2}}{2}\, .
\ee
with 
\begin{equation}
z^{2}= \pi^{2} (Q^{2} P^{2} - (Q.P)^{2}) \, = \pi^{2} \Delta \,  . 
\end{equation}
The variables $(\s, \a, \tau^{a})$ can be regarded as the duality invariant variables. 

\subsection{Conformal compensator, Gauge-fixing, and Analytic Continuation \label{confcomp}}

The constants $C^{I}$ which characterize the localizing instanton solution \eqref{HEKEsol} are all real. Hence, the contour of integration for the variables $s$ and $t$  would appear to be along the real axis. The quadratic terms in $t$ in the  action  \eqref{Sren2} would lead to divergent Gaussian integrals. We will see below that this is nothing but the divergence of Euclidean quantum gravity arising from the integration over the conformal factor that has a wrong sign kinetic term. 

We recall that the scalar kinetic term \eqref{scalarKE2} can be written as 
\be\label{omegaaction}
-\sqrt{-g} g^{\mu\nu}  \left[ e^{-K}K_{IJ} \partial_{\m} X^{I} \partial_{\n} \bar X^{J} - \frac{1}{4}e^{-K}\partial_{\mu} K \partial_{\nu}K  \right]\, .
\ee 
The kinetic term  for the spacetime metric $g_{\mu\nu}$ is of the form\footnote{We  suppress  an overall 
factor of $1/8\pi$ that is irrelevant for the discussion here but is important for the normalization of 
the renormalized action in \S\ref{Macro}.}
\be
 -\frac{1}{6}\sqrt{-g}  e^{-K} R_{g}\, ,
\ee
We can thus identify $e^{-K/2}$ as a conformal compensator $\Omega$ which is often used to extend the gauge principle  to include scale invariance in addition to diffeomorphism invariance. The Einstein-Hilbert action is then replaced by 
\be
 \sqrt{-g}\left[ -\frac{1}{6}  \Omega^{2} \, R_{g}  - g^{\mu\nu }\, \partial_{\mu} \Omega \,\partial_{\nu}\Omega \right] \, ,
\ee
which is  now invariant under both diffeomorphisms and Weyl rescalings. As can be seen from \eqref{omegaaction},  the kinetic term for $\Omega$ has a wrong sign compared to a physical scalar, as is usual for the 
conformal compensator field.  
In D-gauge \cite{Mohaupt:2000mj}
$\Omega$ is gauge-fixed to a constant and one recovers the Einstein-Hilbert action. Our localizing solution is however in a different gauge in which the metric $g$ is gauge-fixed so that $AdS_{2}$ has fixed volume   and hence $\Omega$ is effectively a fluctuating field. 
This also explains why we have $n_{v} +1 $ scalar moduli $\{\phi^{I} \}$ even though there are only $n_{v}$  physical scalars. Essentially, our choice of gauge  enables us to borrow the conformal factor $\Omega$ as an additional scalar degree of freedom. The advantage  is  that the symplectic symmetry acts linearly on the fields $\{ \phi^{I} \}$. 

Since the  kinetic term for conformal compensator  $\Omega$ has a wrong sign, to make the Euclidean functional integral well defined, it is necessary to analytically continue the contour of integration in field space   \cite{Gibbons:1976ue}. 
For our prepotential \eqref{ourprepot}, the K\"ahler potential is given by
\be \label{KahlerpotST}
\exp[-K]  = 4 \, | X^{0}|^{2}\,  \Im (S) \, C_{ab} \, \Im (T^{a}) \, \Im (T^{b}) \, .
\ee
For $S$ and $T^{a}$ fixed, we see that $\Omega$ is proportional to $X^{0}$ up to a  phase that can gauge-fixed by using the additional  $U(1)$ gauge symmetry. Thus, the analytic continuation in the $\Omega$ space can be achieved by analytically continuing in the $X^{0}$ space. For the localizing solution,  $X^{0} =\phi^{0}$. 
Thus,  analytic continuation in $\Omega$ space can be achieved by analytically continuing  in  the  $\phi^{0}$ space. Correspondingly, we take the contour of integration of $\phi^{0}$  or equivalently of  $\s$  along the imaginary axis rather than along the real axis\footnote{In general there can subtleties  in such analytic continuation, see for example  \cite{Harlow:2011ny}. These  will  not be important  in the present context.}.

A familar example of such analytic continuation is the functional integral for the worldsheet metric in first-quantized string theory. The conformal factor of the metric  is the Liouville mode which can be thought of as a conformal compensator. Critical bosonic string with $c=26$  can be regarded as a noncritical string theory with $c=25$ coupled to this Liouville mode. The Liouville mode  plays the role of  time coordinate in target space \cite{Das:1988ds} and has a wrong-sign kinetic term on the worldsheet. The corresponding functional integral then has to be defined by a similar analytic continuation \cite{Polchinski:1998rq}.

\subsection{Evaluation of the Localized Integral \label{Evalu}}

The localizing action $QV$ with abelian gauge fields is purely quadratic. In the gauge that we have chosen the radius of the background $AdS_2 \times S^2$ is set to unity and as a consequence the determinants appear to be independent of charges. However, the physical metric  in Poincaré gravity does depend on charges. Through this dependence,  the over-all normalization of the functional integral is expected to depend on the charges  even if it is independent of the moduli $C^I$ of the localizing instantons. This will contribute to logarithmic corrections to the entropy computed by \cite{Sen:2011ba,Banerjee:2010qc,Banerjee:2011jp}. We do not fully understand the relations between the different gauges to explain this discrepancy.  This would require a careful treatement of the compensating multiplets\footnote{Note that the determinants computed in \cite{Sen:2011ba,Banerjee:2010qc,Banerjee:2011jp} are for the quadratic fluctuation operators obtained by expanding the physical action around the classical black hole background. By contrast, we need determinants for the quadratic fluctuation operators obtained by expanding the localizing action around the localizing instanton.} . Moreover, the overall charge-dependent normalization  is expected to receive contributions also from the fields in the hypermultiplets and gravitini multiplets that we have not taken into account. Our final results and comparisons with the  micrsocopic answer and the macroscopic calculations of \cite{Sen:2011ba,Banerjee:2010qc,Banerjee:2011jp} indicate that  for the $\CN=8$ theory this overall normalization is a constant independent  of charges after including the contributions from all multiplets.

Thus, all that remains is to  compute the determinant of the matrix $M_{IJ}$  introduced in \eqref{Mmat1}. 
Since there are no terms that depend on $\hat A$ for our prepotential, it is homogenous of degree $2$ 
in the variables $X$. As a result, $F_{IJ}X^{J} = F_{I}$, and it follows from \eqref{Ldef} that
\be
K_{I} =   e^{K}N_{IJ} \bar X^{J} \, , \qquad  \bar K_{I} =  e^{K} N_{IJ}  X^{J} \, . 
\ee 
This allows us to write \eqref{Mmat1} as
\be \label{M1}
M_{IJ} = e^{K} \left( N_{IJ} + \frac{1}{4} e^{K} N_{IK} \, p^{K}  N_{JL} \, p^{L}\right) \, .
\ee
We have 
\be
\det (M) =  \exp \left[\frac{(n_{v} + 1 )}{2}K \right] \det (N) \det (1 + \Lambda) \, , 
\ee
where the matrix $\Lambda$ is defined by
\be
\Lambda^{I}_{J} = \frac{1}{4} e^{K} \, p^{I} N_{JL} \, p^{L}  \, . 
\ee
Some elements of this measure such as the  matrix $N_{IJ}$ were anticipated in the work of \cite{LopesCardoso:2006bg, Cardoso:2008fr, Cardoso:2010gc} based on considerations of sympletic invariance. Our derivation follows from the analysis of the induced metric on the localizing manifold and has additional terms depending on $K_{I}$ and $\exp (K)$ which are also sympletic invariant.  Unlike in the $\CN=4$ theory,  in the $\CN=8$ theory  the higher-derivative corrections are zero,  and do not provide  a useful guide for the  determination of nonholomorphic terms of the measure such as the powers of $\exp(K)$. 

It is easy to see that for our system $\textrm{Tr} (\Lambda^{n}) = \lambda^{n}$ where $\lambda$ is a numerical constant independent of charges. As a result, 
\be
\det ( 1+ \Lambda) = \exp (\textrm{Tr} \log ( 1 + \Lambda) ) = \exp ( \log ( 1 + \lambda) ) 
\ee
is a field-independent and charge-independent numerical constant. In what follows, we will ignore 
all such numerical constants in the evaluation of the measure and  determine the overall normalization 
of the functional integral in the end.  

Hence,  up to a constant,   $\det(M)$ is determined by  $\det (N)$ and $\exp (K)$.
For our prepotential,   evaluating on the  localizing instanton solution we obtain
\be
\exp[-K]  = 4 \, P^{2} s = 8 \s/\pi
\ee
which is manifestly duality invariant.
Similarly, 
\be
\det (N )  = \frac{s^{n_{v}-3}  \det (C_{ab})}{4 |X^{0}|^{4}} e^{-2K}  
= s^{n_{v}+3} \,  \big(\frac{P^{2}}{w^{2}} \big)^{2}
\ee
as can be checked using Mathematica. 
In terms of the duality invariant variables defined earlier, we see that  the measure is given by 
\be
\prod_{I=0}^{n_{v}} d\phi^{a} \,\sqrt{\det (N)} = \frac{1}{\sqrt{\s}} \, d\s \, d\a \, \prod_{2}^{n_{v}} d\tau^{a}
\ee
up to an overall constant that is independent of charges and fields. 
The total  measure is thus given by
\be\label{measure1}
\prod_{I=0}^{n_{v}} d\phi^{I} \, \sqrt{\det{(M)}} =  \frac{d\s}{\s^{\rho+ 1}} d\alpha \prod_{2}^{n_{v}} d\tau^{a}\, 
\ee
with $\rho = n_{v}/2$. 
Our total integral is hence manifestly duality invariant.

Performing the Gaussian integrals over $\a$ and $\tau^{a}$ we obtain
\be\label{intrep2}
\int \frac{d\s}{\s^{\rho + 1}} \exp \left( \sigma + \frac{z^{2}}{4\sigma} \, \right) 
\ee
 which  gives exactly the integral representation of  the  Bessel function $\wt I_{7/2}(z)$ for $n_{v} =7$.
The overall numerical  normalization needs to be fixed by hand but once it is fixed for one value of $\Delta$, one  obtains a nontrivial a function for all other values of $\Delta$ given by 
\be
W_{1}(\Delta) =  \sqrt{2} \, \pi \, \big(\frac{\pi}{\Delta} \big)^{7/2} \, I_{7/2}(\pi \sqrt{\Delta}) \, .
\ee
This macroscopic calculation thus precisely reproduces the first term with $c=1$ in \eqref{Wexp} and matches beautifully with  the first term in \eqref{dexp} from the Rademacher expansion  \eqref{rademsp} for  of the microscopic degeneracy $d(\Delta)$.

For large $z$, the Bessel function has an expansion
\begin{equation}\label{Cardy}
I_{\rho} (z) \sim  \frac{e^{z}}{\sqrt{2\pi z}} \left[ 1- \frac{(\mu -1)}{8z} + \frac{(\mu -1)(\mu -3^{2})}{2! (8z)^{3} }- \frac{(\mu -1)(\mu -3^{2}) (\mu -5^{2})}{3! (8z)^{5} } + \ldots \right] \, ,
\end{equation}
with $\mu = 4\rho^{2}$.  The exponential term $\exp (\pi \sqrt{\Delta})$ gives the Cardy formula and $\pi \sqrt{\Delta}$ can be identified with the Wald entropy of the black hole. Higher terms in the series give power-law suppressed finite size corrections to the Wald entropy.  This is however not  a convergent expansion but only an asymptotic expansion. This means that for any given $z$ only the first few terms are useful for making an accurate estimate. Beyond a certain number of terms that depends on a positive power of $z$, including more terms actually makes the estimate worse rather than improve it.  For larger and larger $z$ one can include more or more terms to improve the approximation but this is never convergent for a fixed $z$.

It should be emphasized that  our computation of $W_{1}(\Delta)$ gives an exact integral representation \eqref{intrep2} of the Bessel function $ I_{7/2}(z)$  and not merely the asymptotic expansion \eqref{Cardy}.  This is made possible because localization gives an exact evaluation of the functional integrals and allows one to access large regions in the field space far away from the classical saddle point of the entropy function used to derive the Cardy formula.

It is instructive to compare  the integers $d(\Delta)$ with  the $W_{1}(\Delta)$  and the exponential of the Wald entropy.  We tabulate these numbers  in  Table \eqref{tablefcoeffs2} for the first few values  of $\Delta$. 
\begin{table}[h]   \caption{\small{Comparison of the microscopic degeneracy $d(\Delta)$ with the 
functional integral $W_{1}(\Delta)$ and the exponential of the Wald entropy. 
The last three rows in the table equal each other 
asymptotically.}}   \vspace{8pt}    \centering
   \begin{tabular}{c|cccccccccccccc}   \hline   
       $\Delta$ &  -1 & 0 &3& 4& 7 & 8 &11& 12& 15 \\
       \hline \hline $d(\Delta)$ & 1 & $2$ &8&$12$&39&$56$&152&$208$&513\\  
       \hline $W_{1}(\Delta)$ & 1.040 & $1.855$ &7.972 &$12.201$&38.986&$55.721$&152.041&$208.455$&512.958\\  
        \hline $\exp(\pi\sqrt{\Delta})$ & -  & $1$ &230.765 &$535.492$&4071.93&$7228.35$&33506&$53252$&192401\\   
   \hline    \end{tabular}   \label{tablefcoeffs2}   \end{table}
Note that the area of the horizon goes as $4\pi \sqrt{\Delta}$ in Planck units. Already for $\Delta =12$ this  area would be much larger than one, and one might expect that  the Bekenstein-Hawking-Wald entropy would be a good approximation to the logarithm of the quantum degeneracy. However, we see from the table  that these two differ quite substantially.   Indeed, in this example,  there are no relevant higher-derivative local  terms which arise from integrating out the massive fields. Thus, the Wald entropy equals the Bekenstein-Hawking entropy.  The discrepancy between the degeneracy and the exponential of the Wald entropy arises entirely from  integration over massless fields.   Localization  enables an exact evaluation of these quantum effects. The resulting  $W_{1}(\Delta)$  is in spectacular agreement with $d(\Delta)$ and in fact comes very close to the actual integer even for small values of $\Delta$.  

We see from the asymptotic expansion \eqref{Cardy} that the subleading logarithmic correction to the Bekenstein-Hawking entropy goes as $ -2 \log ({\Delta})$. This in agreement with the results in  \cite{Banerjee:2010qc, Sen:2011ba, Banerjee:2011jp} where the logarithmic correction was computed by evaluating  one-loop determinants of various massless fields around the classical background.  Using localization, this logarithmic correction follows essentially from the analysis of the induced measure on the localizing manifold without the need for any laborious evaluation of one-loop determinants. Moreover, since   localization accesses regions in field space very off-shell from the classical background  the entire series  of power-law suppressed terms in \eqref{Cardy} follows with equal ease.

\subsection{Nonperturbative Corrections, Orbifolds, and Localization\label{Nonpert}}

We have seen that localization
correctly reproduces the first term in the Rademacher expansion. This term already captures all power-law  and logarithmic corrections to the leading Bekenstein-Hawking-Wald entropy exactly to all orders.  
We turn next to the  computation of the higher terms in the Rademacher expansion \eqref{rademsp} with $c >1$. These terms are nonperturbative because they are exponentially suppressed with respect to the terms in  \eqref{Cardy}. 

It was proposed in \cite{Banerjee:2008ky,Sen:2009vz, Murthy:2009dq, Banerjee:2009af} that such non-perturbative corrections could arise from  $\mathbb{Z}_{c}$ orbifolds for all positive integers $c$ because such orbifolds respect the  same boundary conditions  \eqref{asympcond} on the fields.  In general, it is  difficult to justify keeping such subleading exponentials if the  power-law suppressed terms are evaluated only in  an asymptotic expansion. However,  as we have seen, localization gives an exact integral representation of the leading Bessel function in \S\ref{Evalu}.  Since the power-law suppressed contributions are computed exactly,  it is justified to  systematically take into account the exponentially suppressed contributions.

The  $\IZ_{c}$ orbifold configurations that contribute to the localization integral are obtained as follows. We mod out with a symmetry $R_{c}T_{c}$ which   combines a supersymmetric order $c$ twist $R_{c}$ on  $AdS_{2}\times S^{2}$ with an order $c$  shift $T_{c}$ along the $T^{6}$. The orbifold twist  is required to be supersymmetric  because  to preserve the $Q$ supercharge  used for localization, the orbifold action must commute with $L -J$ \cite{Banerjee:2009af}. At the center of  $AdS_{2}$ and at the poles of $S^{2}$ the twist looks like a generator of the supersymmetric $C^{2}/\IZ_{c}$ orbifold.
With an appropriate shift, this action is freely acting and can be used to get smooth solutions \cite{Murthy:2009dq}.  

To illustrate how this works together with localization let us first discuss the case when  $T_{c}(\delta) $ is a simple shift  of $2\pi \delta/c$ along the circle $S^{1}$. It  acts on the momentum modes by
\be\label{phase}
T_{c}(\delta) \, |{m}\rangle = e^{\frac{2\pi i \delta m}{c}} \, | m \rangle \, .
\ee
Let $\phi$ be the azimuthal angle along the $S^{2}$ and $ y$ be the coordinate of the circle $S^{1}$ with $2\pi$ periodicities.  We will denote the orbifolded coordinates with a tilde. The orbifold operaton $R_{c}T_{c}$ identifies points in $AdS_2\times S^2\times S^1$ with the  identification
\begin{equation}\label{identify}
 (\wt \theta, \wt \phi \, , \wt y)\equiv (\wt \theta+\frac{2\pi}{c} \,, \wt  \phi-\frac{2\pi}{c}\, , \wt y+\frac{2\pi \delta}{c})
 \end{equation}
The combined action $R_{c} T_{c} (\delta)$ means that as we go around the boundary of $AdS_{2}$ the momentum modes pick up a phase as in \eqref{phase}. This corresponds to turning on a Wilson line of the Kaluza-Klein gauge field
$\mathcal{A}$  that couples to the  momentum $n$ by modifying  the gauge field as
\be\label{orbA}
\cA =  -i e_{*} (\wt r-1) d \wt\theta  + \delta \, d \wt \theta\
\ee
The  metric on the orbifolded   $AdS_{2}$  factor  has the same form
\be
ds^{2} =  v_{* }\left[  (\wt r^{2} -1)  d\wt \theta^{2}  + \frac{d\wt r^{2}}{(\wt r^{2} -1)} \right] \qquad 1 \leq \wt r < \wt r_{0}; \, \qquad 0 \leq \wt \theta < \frac{2\pi }{c}
\ee 
as the original unorbifolded metric \eqref{metric} but the $\wt \theta$ variable now has a different periodicity and we have cutoff at $\wt r = \wt r_{0}$.   Thus, it is not immediately obvious that asymptotic conditions on the fields are the same  as for the unorbifolded theory. 
To see this, we change coordinates
\begin{equation}
\wt  \theta=\frac{ {\theta}}{c}\, ,\quad \wt \phi= {\phi}-\frac{ {\theta}}{c}\, ,\quad \wt y= {y}+\frac{ {\theta}}{c} \, , \quad \wt r = c r , 
\end{equation}
so that in the  new coordinates, the fields have the same asymptotics \eqref{asympcond}  as before:
\be\label{asymorb}
d s_2^2  \sim  v_{*}  \left[ r^2 d \theta^2 + \frac{d r^2}{r^2} \right] , \qquad 
\cA \sim  -i e_{*} r d \theta\ \, . 
\ee 
Moreover, the new coordinates have the same identification
\begin{equation}
  ( {\theta}, {\phi}, {y})\equiv ( {\theta}+2\pi, {\phi}, {y})\equiv ( {\theta}, {\phi}+2\pi, {y})\equiv ( {\theta}, {\phi}, {y}+2\pi)
\end{equation}
as in the unorbifolded theory.  Such orbifolded field configurations with the same asymptotic behavior will therefore contibute to the functional integral. 

The  orbifold action is freely acting if $\delta$ and $c$ are relatively prime.  Therefore, the localizing equations, which   are local differential equations, remain the same as before and one obtains the same localizing instantons  \eqref{HEKEsol} as before. 
To compute the renormalized action it is convenient to use the tilde coordinates. 
If we  put a cutoff at $r_{0}$, the  range of $r$ is $1/c \leq r \leq r_{0}$ and that of $\wt r$ is   $ 1 \leq \wt r \leq c r_{0}$ . 
 The physical action  is an integral of the same local Lagrangian density as the unorbifolded theory but now the ranges of integration are different.  Since the localizing instantons do not depend on the angular coordinates,  the nontrivial integration is over  the coordinate $\wt r$.  The $r_{0}$ dependent contribution from this integral is therefore $c$ times larger than before but the $r_{0}$ independent constant piece is the same as before.  On the other hand, from the angular integrations  one gets an overall factor of $1/c$ because the range of these coordinates is divided by $c$ by the identification \eqref{identify}.  Altogether, the renormalized action obtained by removing the $r_{0}$ dependent divergence is  smaller by a factor of $c$. Moreover, with the modified gauge field \eqref{orbA} the Wilson line contributes an additional  phase. 
In summary, instead of \eqref{Sren} we obtain 
\be\label{newsolact}
 \exp\left[ \frac{\CS_{ren}(\phi)}{c} +  \,\frac{2 \pi i\,  n \delta}{c} \right] \ ,
\ee
where  $\CS_{ren}$ is the unorbifolded renormalized action for the localizing instantons  given by \eqref{Srenour}. 

Since the phase factor in \eqref{newsolact} does not depend on $\phi$ we can first integrate over $\phi$ as before and  then sum over all phases.  Thus $W_{c}$ factorizes  as
\be
W_{c} (\Delta) = A_{c} (\Delta) B_{c} (\Delta)
\ee
where $A_{c}$  comes from integration over $\phi$ and $B_{c}$  comes from the sum over phases. Since the renormalized action  is now  smaller by a factor of $c$, it is easy to see that the integral $A_{c}$  gives precisely the modified Bessel function but with an argument $z_{c } = z/c$ with possible powers of $c$ coming from the measure which we absorb for now in $B_{c}(\Delta)$.
The final answer thus has the form
\be
W_{c}(\Delta) =  \sqrt{2} \, \pi \, \big(\frac{\pi}{\Delta} \big)^{7/2} \,  
I_{\frac{7}{2}} \big(\frac{\pi \sqrt{\Delta}}{c} \big) \, B_{c}(\Delta) \, .
\ee
This is very close to the $c$-th term in the Rademacher expansion. To obtain agreement we would need to show
\be\label{desired}
B_{c } (\Delta) =  c^{-9/2}K_{c}(\Delta)  \, .
\ee
We see from  \eqref{kloos} that the Kloosterman sum is also a rather intricate sum over various $c$ and $\Delta$ dependent phases. This suggests that by summing over the phases for various allowed orbifolds   and properly  fixing their relative normalization with respect to the $c=1$ term,  it may be possible to compute $B_{c}(\Delta)$ to reproduce the desired expression \eqref{desired} in terms of the Kloosterman sum \cite{Dabholkar:2012}.  

\section{Open Questions and Speculations \label{Open}}

 It is remarkable that a functional integral of string theory in  $AdS_{2}$  precisely reproduces the first term
 in the Rademacher expansion that already captures all power-law suppressed corrections to the Bekenstein-Hawking-Wald formula as described in \S\ref{Evalu}. As we have seen in \S\ref{Nonpert}, the functional integral has all the ingredients to reproduce even the subleading nonperturbative corrections in  the  Rademacher expansion. It would be insteresting to see how the intricate  number theoretic details of the Kloosterman sum \eqref{kloos} will arise from the string theory functional integral \cite{Dabholkar:2012}.  Since $d(\Delta)$ is an integer, $W(\Delta)$ would also have to be an integer. This suggests an underlying integral structure in  quantum gravity at a deeper level.
 
Our  computation suggests that the bulk $AdS$ string theory is every bit as fundamental as the boundary $CFT$. Even though one sometimes refers to the $AdS$ computation as macroscopic and thermodynamic,  quantum gravity in $AdS_{2}$ does  not appear to be an emergent, coarse-grained description of the more microscopic boundary theory. Each theory has its own   rules of computation. It seems  more natural to regard $AdS/CFT$ holography as an exact strong-weak coupling duality.
 
So far we have used holography in its original sense to mean a complete accounting of the degrees of freedom associated with the  $AdS_{2}$ black hole horizon in terms of  the states of a $CFT_{1}$ in one lower dimension.  The  $AdS_{2}/CFT_{1}$ correspondence actually extends this idea further to apply correlation functions as well. The boundary $CFT_{1}$  has a  $GL(d)$ symmetry that acts upon $d(q, p)$ zero energy states.  The observables of the theory are thus simply $d \times d$ matrices $\{M_{i}\}$.  A precise state-operator correspondence has been suggested \cite{Sen:2011cn} that allows one to define, at least formally, the corresponding correlation functions for some of the observables in the bulk theory.
 In the boundary theory it is easy to define correlation functions of 
observables as traces of strings of operators such as 
\begin{equation}
\textrm{Tr} (M_{1}M_{2}\ldots M_{k}) \ .
\end{equation}
We have seen that   localization techniques can be successfully applied for computing the partition function to compute the integer $d$. A natural question is if localization can be useful for computing the  correlation functions such as above. Such a computation would allow us to  recover the discrete information  about the microstates of a black hole from observables living in the bulk near the horizon. This of course goes to heart of the problem of information retrieval from black holes. It is likely that one would need to extend the localization analysis beyond the massless fields to higher string modes to access this information.

The content of the boundary $CFT_{1}$ is essentially completely determined by the integer $d$. The  bulk theory  has an elaborate field content and action that depends on the compactification $K$ and the charges of the black hole. Imagine two different bulk theories $AdS_{2}\times K$ and $AdS_{2} \times K'$  but with the same black hole degeneracy  $d$. This would suggest 
that the two string theories near the horizon of two  very different black holes in  very different compactifications are dual to the same $CFT_{1}$. 
 By transitivity of duality, this would imply that the two string theories themselves are dual to each other. This  conclusion seems inescapable  from the perspective of the $CFT_{1}$.  Note that it is not easy to arrange the situation when the degeneracies of two different black holes are given by the same integer.  For example, if the degeneracy is given by the Fourier coefficients of some modular form, it would be  rare, but not impossible,  that two such Fourier coefficients are precisely equal.

Our analysis uses an $\CN=2$ reduction of the full $\CN=8$ theory by dropping six gravitini multiplets of $\CN=2$ and the hypermultiplets. This is partially motivated by the fact that the hypermultiplets are flat directions of the classical entropy function and our black hole is not charged under the gauge fields that belong to the gravitini multiplets. We have also ignored  D-terms. This is partially justified by the fact that the black hole horizon is supersymmetric and a large class of D-terms are known not to contribute to the Wald entropy as a consequence of this supersymmetry \cite{deWit:2010za}. 
Our final answer strongly suggests that these assumptions are justified and our reduced theory fully captures the physics. A technical obstacle in analyzing the validity of these assumptions stems from the fact that the incorporation of  the hypermultiplets and the gravitini multiplets  would require infinite number of auxiliary fields if all $\CN=8$ supersymmetries are realized off-shell. It may be possible to make progress in this direction  perhaps by using a formulation where only the Q-supersymmetry used for localization is realized off-shell but on all fields of $\CN=8$ supergravity. Alternatively, it may be possible to repeat the localization analysis in a different off-shell formalism such as the harmonic superspace \cite{Galperin:2001uw}
where all $\CN=8$ supersymmetries are realized off-shell with infinite number of auxiliary fields; but perhaps only a small number of auxiliary fields get excited for the localizing solution.

\subsection*{Acknowledgments}

It is a pleasure to thank  Ashoke Sen and Edward Witten for useful  discussions. The work of A.~D. was
supported in part by the Excellence Chair of the Agence Nationale de la
Recherche (ANR).
The work of J.~G. was supported in part by Fundac\~{a}o para Ci\^{e}ncia e
Tecnologia (FCT). The work of S.~M. was supported by the European 
Commission Marie Curie Fellowship under the contract PIIF-GA-2008-220899, and 
is presently supported by the ERC Advanced Grant no. 246974,
{\it ``Supersymmetry: a window to non-perturbative physics''}.
A.~D. would like acknowledge the hospitality at the Benasque Center.  J.~G. would like acknowledge hospitality of the HRI and TIFR where part of this work was completed. This  material is based  upon work supported in part by the National Science Foundation under Grant No. 1066293 and the hospitality of the Aspen Center for Physics.
%

\providecommand{\href}[2]{#2}\begingroup\raggedright\endgroup

\end{document}